\documentclass[useAMS,usenatbib]{mn2e}

\usepackage{amsfonts,amsmath,amssymb,bm,epsfig}

\let\ph\varphi
\let\th\theta
\newcommand{\vecnab}{\bm{\nabla}}
\newcommand{\ud}{\mathrm{d}}

\allowdisplaybreaks

\title[MOND effects in the Solar system]{External field effect of modified Newtonian dynamics \\in the Solar system}

\author[Luc Blanchet \& J\'er\^ome Novak]{Luc Blanchet,$^1$\thanks{E-mail: blanchet@iap.fr}
  J\'erome Novak$^2$\thanks{E-mail: jerome.novak@obspm.fr}\\
  $^1\mathcal{G}\mathbb{R}\varepsilon{\mathbb{C}}\mathcal{O}$,
  Institut d'Astrophysique de Paris --- UMR 7095 du CNRS, Universit\'e Pierre
  et Marie Curie,\\ 98$^{\rm bis}$ boulevard Arago, 75014 Paris, France\\
  $^2$Laboratoire Univers et Th\'eories, Observatoire de
  Paris, CNRS, Universit\'e Paris Diderot,\\ 5 place Jules Janssen,
  92190 Meudon, France}
\date{\today}

\pagerange{\pageref{firstpage}--\pageref{lastpage}} \pubyear{2010}

\begin{document}
\label{firstpage}

\maketitle

\begin{abstract}
The Modified Newtonian Dynamics (MOND) has been formulated as a modification of the Poisson equation for the Newtonian gravitational field. This theory generically predicts a violation of the strong version of the equivalence principle, and as a result the gravitational dynamics of a system depends on the external gravitational field in which the system is embedded. This so-called external field effect has been recently shown to imply the existence of an anomalous quadrupolar correction, along the direction of the external galactic field, in the gravitational potential felt by planets in the Solar System. In this paper we confirm the existence of this effect by a numerical integration of the MOND equation in the presence of an external field, and compute the secular precession of the perihelion of planets induced by this effect. We find that the precession effect is rather large for outer gaseous planets, and in the case of Saturn is comparable to published residuals of precession obtained by Saturn range tracking data. The effect is much smaller for inner planets, but in the case of the Earth it appears to be in conflict for most of the MOND functions $\mu(y)$ with the very good constraint on the perihelion precession obtained from Jupiter VLBI data. The MOND functions that are compatible with this constraint appear to have a very rapid transition from the MONDian regime to the Newtonian one. 
\end{abstract}


\section{Introduction} 
\label{s:intro}

\subsection{Motivation}
\label{ss:motiv}

The Modified Newtonian Dynamics (MOND) --\cite{Milg1,Milg2,Milg3}-- is a successful alternative framework for interpreting the galactic rotation curves and the empirical Tully-Fisher relation without relying on dark matter halos (see~\citealt{SandMcG02} for a review). At the non-relativistic level, the best modified-gravity formulation of MOND is the modified Poisson equation originally proposed by \cite{BekM84},
\begin{equation}
  \label{e:MOND}
\vecnab \cdot \left[ \mu\left(\frac{g}{a_0}\right) \vecnab U
  \right] = -4\pi G \rho\,,  
\end{equation}
where $\rho$ is the density of ordinary (baryonic) matter, $U$ is the gravitational potential, $\bm{g}=\vecnab U$ is the gravitational field and $g = \vert\bm{g}\vert$ its ordinary Euclidean norm. The modification of the Poisson equation is encoded in the MOND function $\mu(y)$ of the single argument $y\equiv g/a_0$, where $a_0=1.2\times 10^{-10}\,\mathrm{m}/\mathrm{s}^2$ denotes the MOND constant acceleration scale. The MOND function interpolates between the MOND regime corresponding to weak gravitational fields $y=g/a_0\ll 1$, for which it behaves as $\mu(y)=y+o(y)$, and the Newtonian strong-field regime $y\gg 1$, where $\mu$ reduces to $1$ so that we recover the usual Newtonian gravity.

Relativistic extensions of MOND modifying general relativity have been proposed \citep{Bek04,Sand05} and extensively studied (see~\citealt{Bekrev} for a review). Alternatively the MOND equation~\eqref{e:MOND} can be reinterpreted as a modification of dark matter rather than gravity by invoking a mechanism of ``gravitational polarisation'' --- a gravitational analogue of the electric polarisation \citep{B07mond}; this yields to the concept of dipolar dark matter which has been formulated as a relativistic model in standard general relativity by \cite{BL08,BL09}.

An important consequence of the non-linearity of Eq.~\eqref{e:MOND} in the MOND regime, is that the gravitational dynamics of a system is influenced (besides the well-known tidal force) by the external gravitational environment in which the system is embedded. This is known as the external field effect (EFE), which has non-trivial implications for non-isolated gravitating systems. The EFE was conjectured to explain the dynamics of open star clusters in our galaxy \citep{Milg1,Milg2,Milg3}, since they do not show evidence of dark matter despite the involved weak internal accelerations (i.e. below $a_0$). The EFE effect shows that the dynamics of these systems should actually be Newtonian as a result of their immersion in the gravitational field of the Milky Way. The EFE is a rigorous prediction of the Bekenstein-Milgrom equation \eqref{e:MOND}, and is best exemplified by the asymptotic behaviour of the solution of \eqref{e:MOND} far from a localised matter distribution (say, the Solar System), in the presence of a constant external gravitational field $\bm{g}_\text{e}$ (the field of the Milky Way). At large distances $r=\vert\mathbf{x}\vert\to\infty$ we have \citep{BekM84}
\begin{equation}
  \label{e:asym}
  U = \bm{g}_\text{e}\cdot\mathbf{x} + \frac{GM/\mu_\text{e}}{r \sqrt{1 + \lambda_\text{e} \sin^2 \th}} + \mathcal{O} \left(
    \frac{1}{r^2} \right)\,,
\end{equation}
where $M$ is the mass of the localised matter distribution, where $\th$ is the azimuthal angle from the direction of the external field $\bm{g}_\text{e}$ (see also Fig.~\ref{f:setting}), and where we denote $\mu_\text{e} \equiv \mu(y_\text{e})$ and $\lambda_\text{e} \equiv y_\text{e} \mu'_\text{e}/\mu_\text{e}$, with $y_\text{e}=g_\text{e}/a_0$ and $\mu'_\text{e}= \ud\mu(y_\text{e})/\ud y_\text{e}$. In the presence of the external field, the MOND internal potential $u\equiv U-\bm{g}_\text{e}\cdot\mathbf{x}$ shows a Newtonian-like fall-off $\sim r^{-1}$ at large distances but with an effective gravitational constant $G/\mu_\text{e}$.\footnote{Recall that in the absence of the external field the MOND potential behaves like $U\sim -\sqrt{GM a_0}\ln r$, showing that there is no escape velocity from an isolated system \citep{FBZ08}. However since no object is truly isolated the asymptotic behaviour of the potential is always given by \eqref{e:asym}, in the approximation where the external field is constant.} However, contrary to the Newtonian case, it exhibits a non-spherical deformation along the direction of the external field. The fact that the external field $\bm{g}_\text{e}$ does not disappear from the internal dynamics can be interpreted as a violation of the strong version of the equivalence principle.\footnote{However the weak version of the equivalence principle, that all test particles in the MOND gravitational field have universal acceleration $\mathbf{a}=\bm{g}$, remains satisfied \citep{BekM84}.} For the reader's convenience we derive the result \eqref{e:asym} in the Appendix~\ref{appA}.

In a recent paper, \cite{Milg09} has shown that the imprint of the external galactic field $\bm{g}_\text{e}$ on the Solar System (due to a violation of the strong equivalence principle) shows up not only asymptotically, but also in the inner regions of the system, where it may have implications for the motion of planets. This is somewhat unexpected because gravity is strong there (we have $g\gg a_0$) and the dynamics should be Newtonian. However, because of the special properties of the equation \eqref{e:MOND}, the solution will be given by some modified Poisson-type integral, and the dynamics in the strong-field region will be affected by the anomalous behaviour in the asymptotic weak-field region. Thus the results apply only to the nonlinear Poisson equation \eqref{e:MOND}, not to modified inertia formulations of MOND. The anomaly expresses itself as an anomalous quadrupolar contribution to the MONDian internal field $u$, as compared to the Newtonian potential, given by
\begin{equation}\label{deltaU}
\delta u = \frac{1}{2} x^ix^j Q_{ij}\,,
\end{equation}
where $x^i$ is the distance from the Solar System barycentre, and $Q_{ij}$ is a trace-free quadrupole moment of the type
\begin{equation}\label{Qij}
Q_{ij}=Q_2\Bigl(e^ie^j-\frac{1}{3}\delta_{ij}\Bigr)\,,
\end{equation}
with $Q_2$ being the quadrupole coefficient and $\bm{e}=(e^i)$ the preferred direction of the galactic centre (i.e. $\bm{e}=\bm{g}_\text{e}/g_\text{e}$). The anomalous term \eqref{deltaU} is a harmonic solution of the Laplace equation, i.e. $\Delta\delta u = 0$. 

The radial dependence of this anomaly is $\propto r^2$ and can thus be separated from a quadrupolar deformation due to the Sun's oblateness which decreases like $\propto r^{-3}$. This result is valid whenever $r$ is much less than the MOND transition distance for the Solar System, defined by $r_0=\sqrt{G M/a_0}$ with $M$ the mass of the Sun and $a_0$ the MOND acceleration scale. This radius corresponds to the transition region where the Newtonian acceleration becomes of the order of the MOND acceleration $a_0$ and therefore, MOND effects become dominant. We have $r_0\simeq 7100\,\mathrm{AU}$ so the effect \eqref{deltaU}--\eqref{Qij} is valid in a large volume around the Sun including all the planets (recall that Neptune's orbit is at $30\,\mathrm{AU}$).

In addition to \eqref{deltaU} we have also the usual MOND effect. In the special case of spherical symmetry, the equation \eqref{e:MOND} reduces to $\mu(g/a_0)g=g_\text{N}$, with $g_\text{N}$ the ordinary Newtonian gravitational field. For a MOND function behaving like $\mu(y)\sim 1-\epsilon y^{-\gamma}$ when $y\to\infty$ (i.e. in the region $r\ll r_0$), the Newtonian field is modified by $\delta g_\text{N}\simeq\epsilon a_0(r/r_0)^{2\gamma-2}$. For $\gamma=1$ this corresponds to a constant acceleration similar to the Pioneer anomaly, and for $\gamma\geqslant 2$ the effect is very small and the motion of planets is almost unaffected. This effect is independent of the presence of the external field, and is spherical, so can be distinguished from the quadrupolar deformation \eqref{deltaU} induced by the external field; we neglect this effect here.

\subsection{Summary}
\label{ss:sum}

In the present paper we shall confirm the existence of the effect \eqref{deltaU}--\eqref{Qij} and shall derive it by an independent numerical integration of the equation~\eqref{e:MOND} using an elliptic solver originally built for numerical relativity purposes.\footnote{The code is based on the library \textsc{lorene}, available from the website \texttt{http://www.lorene.obspm.fr}.} The magnitude of the quadrupole coefficient $Q_2$ depends on the dimensionless ratio between the external field $g_\text{e}$ and the MOND acceleration $a_0$, say
\begin{equation}
  \label{eta}
  \eta = \frac{g_\text{e}}{a_0}\,,
\end{equation}
and on the particular MOND interpolating function $\mu$ in use. For the Milky Way field at the level of the Sun we have $g_\text{e}\simeq 1.9\times 10^{-10}\,\mathrm{m}/\mathrm{s}^2$ which happens to be slightly above the MOND scale, i.e. $\eta \simeq 1.6$. Our calculation implemented for various standard choices for the $\mu$-function gives
\begin{equation}
  \label{e:interv_Qintro}
  2.1\times 10^{-27} \textrm{ s}^{-2} \lesssim Q_2 \lesssim 4.1\times
10^{-26} \textrm{ s}^{-2}\,.
\end{equation}
The quadrupole coefficient $Q_2$ is found to be positive corresponding to a prolate elongation along the quadrupolar axis. We shall verify numerically that $Q_2$ is indeed constant over a large portion of the inner Solar System, and starts decreasing when the distance becomes comparable to the MOND transition scale $r_0$, at a few thousands AU from the Sun. We shall also check that $Q_2$ scales approximately like the inverse square root of the central mass, in agreement with dimensional analysis which shows that we should approximately have $Q_2 = q_2\,a_0/r_0$, where $q_2(\eta)$ is a dimensionless quadrupole coefficient depending on the ratio $\eta$ between the external field and the MOND acceleration. We find that for $\eta \simeq 1.6$, $q_2$ ranges over $0.02 \lesssim q_2 \lesssim 0.36$, in good agreement with the values found by \cite{Milg09}. 

We then study some consequences of the anomalous quadrupole moment for the motion of planets, notably the secular advances of planetary perihelia. Applying standard perturbation equations of celestial mechanics we obtain the anomalous precession rate 
\begin{equation}\label{Delta}
\Delta_2 = \left\langle\frac{\ud \tilde{\omega}}{\ud t}\right\rangle = \frac{Q_2 \sqrt{1-e^2}}{4 n}\Bigl[ 1 + 5 \cos (2\tilde{\omega})\Bigr]\,.
\end{equation}
To simplify, we have assumed that the direction of the galactic centre lies in the plane of the ecliptic (this is only approximately correct) and that all planets move in this plane. We defined the origin of the precession angle $\tilde{\omega}$ to be the direction of the galactic centre. The orbital frequency of the planet is $n=2\pi/P$ ($P$ is the orbital period), and $a$, $e$ and $\tilde{\omega}=\omega+\Omega$ denote the standard orbital elements. The effect seems to be the most interesting for Saturn for which we find
\begin{equation}
  \label{e:interv_Delta}
  0.3 \,\textrm{mas}/\textrm{cy}\lesssim \Delta_2 \lesssim 5.8 \,\textrm{mas}/\textrm{cy}.
\end{equation}
Those values are within the range of published residuals for the precession of Saturn as obtained from global fits of the Solar System dynamics in \cite{pitjeva05} and \cite{fienga09}. However we shall find that in the case of the Earth the predicted perihelion precession may be greatly constrained by the best estimates of postfit residuals obtained thanks to the Jupiter VLBI data using the INPOP planetary ephemerides \citep{fienga09}. This is turn reduces the possible choices for the MOND function $\mu(y)$ to those that exhibit a sharp transition from the Newtonian to MONDian regime.

To conclude, we present very accurate numerical computations performed within the well-defined theory \eqref{e:MOND}, and compare the results with recent observations of the motion of planets in the Solar System. We find that the observational constraints are rather strong, and may even conflict with some of the predictions. Thus our results provide motivation for investigating more systematically possible anomalous effects in the Solar System predicted by alternative theories. In particular, the external field effect associated with the preferred direction of the galactic centre could be seen as a typical one arising in generic attempts to modify gravity with motivation coming from dark matter and/or dark energy. Indeed we expect that generic departures from general relativity will violate the strong equivalence principle \citep{Wthexp}.

On the other hand let us cautiously remark that MOND and more sophisticated theories such as TeVeS \citep{Bek04}, which are intended to describe the weak field regime of gravity (below $a_0$), may not be extrapolated without modification to the strong field of the Solar System. For instance it has been argued by \cite{FB05} that a MOND interpolating function $\mu$ which performs well at fitting the rotation curves of galaxies is given by $\mu_1$ defined by \eqref{mu1} below. However this function has a rather slow transition to the Newtonian regime, given by $\mu_1\sim 1-y^{-1}$ when $y=g/a_0\to\infty$, which is already excluded by Solar System observations \citep{sereno-06}. Indeed such slow fall-off $-y^{-1}$ predicts a constant supplementary acceleration directed toward the Sun $\delta g_\text{N}=a_0$ (i.e. a ``Pioneer'' effect), which is ruled out because not seen from the motion of planets. Thus it could be that the transition between MOND and the Newtonian regime is more complicated than is modelled in Eq.~\eqref{e:MOND}. This is also true for the modified dark matter model \citep{BL08,BL09} which may only give an effective description valid in the weak field limit and cannot be extrapolated as it stands to the Solar System. While looking at MOND-like effects in the Solar System we should keep the previous \textit{proviso} in mind. The potential conflict we find here with the Solar System dynamics (notably with the Earth orbital precession) may not necessarily invalidate those theories if they are not ``fundamental'' theories but rather ``phenomenological'' models. 

The plan of this paper is the following. In Sec.~\ref{s:multipoles}, we derive an expression of the solution of the MOND equation near the origin in terms of multipole moments. Section~\ref{s:numerics} is devoted to numerical techniques and results, with a description of the particular formulation and assumptions used for the numerical integration in Sec.~\ref{ss:implementation}, of the various tests passed by the code (Sec.~\ref{ss:tests}), and the numerical results and values for the first multipoles of the potential in Sec.~\ref{ss:quadrupole}. Section~\ref{s:effect} details the consequences of this modified gravitational potential on the orbits of Solar-system planets, with first the derivation of the perturbation equations in Sec.~\ref{ss:perturbation} and numerical values for the planetary precessions (Sec.~\ref{ss:precession}). Finally, Section~\ref{s:conc} summarises the results and gives some concluding remarks.

\section{Multipolar expansion of the MONDian potential}
\label{s:multipoles}

In this paper we shall solve the modified Poisson equation \eqref{e:MOND} with the boundary condition given by the constant external gravitational field $\bm{g}_\text{e}$ (defining a preferred spatial direction denoted $\bm{e}=\bm{g}_\text{e}/g_\text{e}$), i.e. that the gravitational field $\bm{g}=\vecnab U$ asymptotes to $\lim_{r\to \infty} \bm{g} = \bm{g}_\text{e}$. The external field $\bm{g}_\text{e}$ should consistently obey a MOND equation, but here we shall simply need to assume that $\bm{g}_\text{e}$ is constant over the entire Solar System. The MONDian physicist measures from the motion of planets relatively to the Sun the internal gravitational potential $u$ defined by
\begin{equation}
  \label{e:def_u}
  u = U - \bm{g}_\text{e} \cdot \mathbf{x}\,,
\end{equation}
which is such that $\lim_{r\to \infty}u = 0$. Contrary to what happens in the Newtonian case, the external field $\bm{g}_\text{e}$ does not disappear from the gravitational field equation \eqref{e:MOND} and we want to investigate numerically its effect. The anomaly detected by a Newtonian physicist with respect to the MONDian physicist is the difference of internal potentials,
\begin{equation}
  \label{uuN}
\delta u = u - u_\text{N}\,,
\end{equation}
where $u_\mathrm{N}$ denotes the ordinary Newtonian potential generated by the same ordinary matter distribution $\rho$, and thus solution of the Poisson equation $\Delta u_\text{N}= -4\pi G \rho$ with the boundary condition that $\lim_{r\to \infty}u_\text{N} = 0$. Hence $u_\mathrm{N}$ is given by the standard Poisson integral
\begin{equation}
  \label{deltauN}
u_\mathrm{N}(\mathbf{x},t) = G \int \frac{\ud^3 \mathbf{x}'}{\vert\mathbf{x}-\mathbf{x}'\vert}\,\rho(\mathbf{x}',t)\,.
\end{equation}
A short calculation shows that the anomaly obeys the ordinary Poisson equation $\Delta \delta u = -4\pi G \rho_\text{pdm}$, where $\rho_\text{pdm}$ is the density of ``phantom dark matter'' defined by
\begin{equation}
  \label{pdm}
\rho_\text{pdm} = \frac{1}{4\pi G}\vecnab\cdot\left(\chi\vecnab U\right)\,,
\end{equation}
where we denote $\chi\equiv\mu-1$. The phantom dark matter represents the mass density that a Newtonian physicist would attribute to dark matter. In the model by \cite{BL08,BL09} the phantom dark matter is interpreted as the density of polarisation of some dipolar dark matter medium and the coefficient $\chi$ represents the ``gravitational susceptibility'' of this dark matter medium.

The Poisson equation $\Delta \delta u = -4\pi G \rho_\text{pdm}$ is to be solved with the boundary condition that $\lim_{r\to \infty}\delta u = 0$; hence the solution is given by the Poisson integral
\begin{equation}
  \label{deltaupoisson}
\delta u(\mathbf{x},t) = G \int \frac{\ud^3 \mathbf{x}'}{\vert\mathbf{x}-\mathbf{x}'\vert}\,\rho_\text{pdm}(\mathbf{x}',t)\,.
\end{equation}
We emphasise here that, contrary to the Newtonian (linear) case, the knowledge of the matter density distribution does not allow to obtain any analytic solution for the potential; however, we can still infer the structure of the multipolar expansion near the origin, and the moments will be computed numerically. We can check that the phantom dark matter behaves like $r^{-3}$ when $r\to \infty$, so the integral \eqref{deltaupoisson} is perfectly convergent.

We want to discuss the influence of the external galactic field in the inner part of the Solar System where the gravitational field is strong ($g\gg a_0$); thus $\mu$ tends to one there, so $\chi$ tends to zero. For the discussion in this section we adopt the extreme case where $\chi$ is \textit{exactly} zero in a neighbourhood of the origin, say for $r\leqslant\varepsilon$, so that there is no phantom dark matter for $r\leqslant\varepsilon$; in later sections devoted to the full numerical integration we shall still make this assumption by posing $\chi=0$ inside the Sun (in particular we shall always neglect the small MOND effect at the centre of the Sun where gravity is vanishingly small). If $\rho_\text{pdm}=0$ when $r\leqslant\varepsilon$ we can directly obtain the multipolar expansion of the anomalous term \eqref{deltaupoisson} about the origin by Taylor expanding the integrand when $r=\vert\mathbf{x}\vert\to 0$. In this way we obtain
\begin{equation}
  \label{multexp}
\delta u = \sum_{l=0}^{+\infty} \frac{(-)^l}{l!} \,x^L Q_L\,,
\end{equation}
where the multipole moments near the origin are given by\footnote{Our notation is as follows: $L = i_1 \cdots i_l$ denotes a multi-index composed of $l$ multipolar spatial indices $i_1, \cdots, i_l$ (ranging from 1 to 3); $\partial_L = \partial_{i_1} \cdots \partial_{i_l}$ is the product of $l$ partial derivatives $\partial_i \equiv \partial / \partial x^i$; $x^L = x^{i_1} \cdots x^{i_l}$ is the product of $l$ spatial positions $x^i$; similarly $n^L = n^{i_1} \cdots n^{i_l} = x^L/r^l$ is the product of $l$ unit vectors $n^i=x^i/r$; the symmetric-trace-free (STF) projection is indicated with a hat, for instance $\hat{x}^L \equiv \text{STF}[x^L]$, and similarly for $\hat{n}^L$ and $\hat{\partial}_L$. The decomposition of $\partial_L$ in terms of STF components $\hat{\partial}_L$ is given by \eqref{STFformula}--\eqref{alk} below. In the case of summed-up (dummy) multi-indices $L$, we do not write the $l$ summations from 1 to 3 over their indices.} 
\begin{equation}
  \label{QLpdm}
Q_L = G \int_{r > \varepsilon} \ud^3 \mathbf{x} \,\rho_\text{pdm}\,\partial_L\left(\frac{1}{r}\right)\,,
\end{equation}
Because the integration in \eqref{QLpdm} is limited to the domain $r>\varepsilon$ and $\partial_L(1/r)$ is symmetric-trace-free (STF) there [indeed $\Delta(1/r)=0$], we deduce that the multipole moments $Q_L$ themselves are STF. This can also be immediately inferred from the fact that $\Delta \delta u = 0$ when $r\leqslant\varepsilon$, hence the multipole expansion \eqref{multexp} must be a homogeneous solution of the Laplace equation which is regular at the origin, and is therefore necessarily made solely of STF tensors of type $\hat{x}^L$. Hence we can replace $x^L$ in \eqref{multexp} by its STF projection $\hat{x}^L$. It is clear from the formula \eqref{QLpdm} that the MONDian gravitational field (for $r \geqslant r_0$) can influence the near-zone expansion of the field when $r \to 0$.

With the expression \eqref{pdm} for the phantom dark matter and the MOND equation \eqref{e:MOND}, we can further transform $Q_L$ as
\begin{equation}
  \label{QL}
Q_L = -\frac{1}{4\pi} \int_{r > \varepsilon} \ud^3 \mathbf{x}\Bigl[4\pi G \rho + \Delta U\Bigr] \partial_L\left(\frac{1}{r}\right)\,.
\end{equation}
Approximating the central matter distribution as being spherically symmetric (i.e. ignoring the ``back-reaction'' of the non-spherical anomalous field $\delta u$ on the matter which generates the field), we see that the first term is non-zero only in the monopolar case $l=0$ where it reduces in the limit $\varepsilon\to 0$ to minus the Newtonian potential evaluated at the origin. On the other hand the second term in \eqref{QL} can be transformed as a surface integral. We find
\begin{equation}
  \label{QLsurf}
Q_L = - u_\text{N}(\mathbf{0})\delta_{l,0} + \frac{1}{4\pi} \int_{r=\varepsilon}^{r=\infty} \ud^2 S_i \Bigl[ U \partial_{iL}\left(\frac{1}{r}\right) - \partial_i U \partial_{L}\left(\frac{1}{r}\right)\Bigr]\,.
\end{equation}
Our notation means that the surface integral is composed of two contributions, an inner one at $r=\varepsilon$ (denoted $Q_L^\varepsilon$) and an outer one at infinity $r=\infty$ (say $Q_L^\infty$). In Eq.~\eqref{QLsurf} we implicitly assume that the limit $\varepsilon\to 0$ is to be taken after evaluating the inner surface integral.

Inserting $U=u+\bm{g}_\text{e} \cdot \mathbf{x}$ into the surface integral at infinity $Q_L^\infty$ we find that it contributes only to the dipolar term and reduces to the external galactic field; thus $Q_i^\infty=g_\text{e}^i$ with all other $Q_L^\infty$'s being zero. On the other hand the inner surface integral $Q_L^\varepsilon$ is found to have a well-defined limit when $\varepsilon\to 0$ given by $Q_L^\varepsilon=(-)^l (\hat{\partial}_L U)(\mathbf{0})$, where we recall that $\hat{\partial}_L$ denotes the STF part of $\partial_L = \partial_{i_1} \cdots \partial_{i_l}$. We then find that the galactic field in $U=u+\bm{g}_\text{e} \cdot \mathbf{x}$ cancels the dipole term in the surface integral at infinity, so that the result is 
\begin{equation}
  \label{QLres}
Q_L = - u_\text{N}(\mathbf{0})\delta_{l,0} + (-)^l (\hat{\partial}_L u)(\mathbf{0})\,.
\end{equation}
The internal MONDian potential $u$ admits therefore the following STF multipolar expansion (equivalent to a near-zone expansion when $r\to 0$)
\begin{equation}
  \label{expu}
u = u_\text{N} - u_\text{N}(\mathbf{0}) + \sum_{l=0}^{+\infty} \frac{1}{l!} x^L (\hat{\partial}_L u)(\mathbf{0})\,.
\end{equation}
In Appendix~\ref{appB} we derive an alternative proof of this result. At this stage we have elucidated the structure of the multipole expansion of the anomaly $\delta u$ near the origin, but still we need to resort to a numerical integration of the non-linear MOND equation in order to obtain quantitative values for the multipole moments. Section \ref{s:numerics} will be devoted to this task.

Finally we show that the dipole moment $Q_i$ (with $l=1$) is actually zero \citep{Milg09}. This follows from the weak equivalence principle satisfied by the Bekenstein-Milgrom theory. As a consequence, the total acceleration of the centre of mass of the Solar System in the galactic external field $\bm{g}_\text{e}$ does not depend on the Solar System's internal dynamics and is simply given by $\bm{g}_\text{e}$ (for a detailed proof see~\citealt{BekM84}). The centre of mass of the Solar System does not differ much from that of the Sun, so we deduce that the ``self-acceleration'' of the Sun, i.e. the total acceleration due to the internal potential $u$ when integrated over the whole volume of the Sun, must be (approximately) zero:
\begin{equation}
  \label{deltau3_int}
\int \ud^3 \mathbf{x}\,\rho\,\partial_i u = 0\,.
\end{equation}
We have obviously the same result for the Newtonian potential $u_\text{N}$ in Newtonian gravity so the same will be true for the anomalous field $\delta u = u - u_\text{N}$. In other words the phantom dark matter which is the source for the anomaly exerts no net force on the Sun and we get, for a spherically symmetric star, 
\begin{equation}
  \label{deltau3_spher}
Q_i = -\frac{1}{M}\int \ud^3 \mathbf{x}\,\rho\,\partial_i \delta u = 0\,.
\end{equation}
In Sec.~\ref{ss:quadrupole} we shall numerically verify that the dipole moment $Q_i$ is indeed zero within our numerical error bars.

\section{Numerical integration of the MOND equation}
\label{s:numerics}

\subsection{Formulation and implementation}
\label{ss:implementation}

\begin{figure}
  \centerline{\includegraphics[width = 8cm]{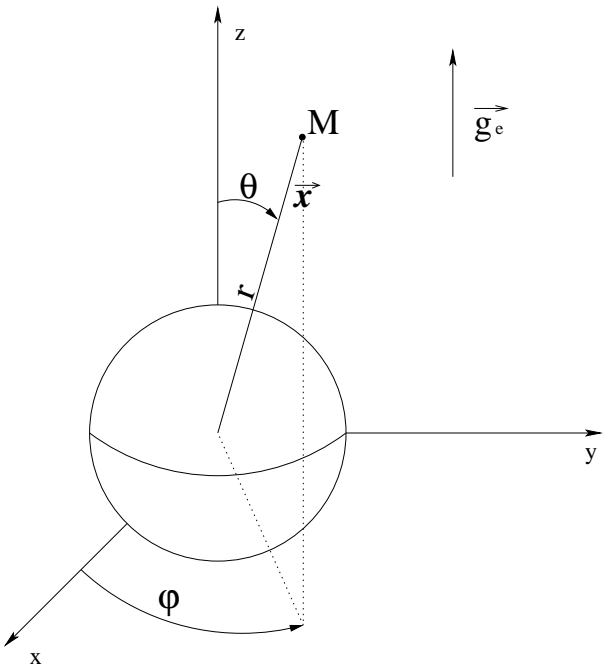}}
  \caption{Setting of the spherically symmetric star and the asymptotic
    galactic field $\bm{g}_{\text{e}}$, using spherical coordinates $\{r, \th, \ph\}$.}
  \label{f:setting}
\end{figure}

Using spherical spatial coordinates $\{r, \th, \ph\}$, we consider a star represented by a spherically symmetric distribution of matter $\rho(r)$ obtained from a hydrostatic equilibrium model in Newtonian theory (polytrope). These models are obtained by integrating the hydrostatic spherical equilibrium equation (relating the pressure $p$ and the gravitational field) and the Newtonian equation for the gravitational field (relating the gravitational field and the density distribution $\rho$), together with a polytropic equation of state (EOS) of the form $p=K\rho^\gamma$, where $K$ and $\gamma$ are two constants. As it shall be shown later in Sec.~\ref{ss:tests}, the results do not depend on the specific EOS used to obtain this hydrostatic equilibrium, so we do not discuss the particular EOS used for obtaining the density distribution. It also means that we neglect the effect of MOND theory on the structure of the star. This is not a severe restriction, since inside the star the gravitational field is much higher in amplitude than the constant $a_0$, making $\mu \sim 1$ and Newtonian theory is recovered up to very high accuracy. We then solve the MOND equation~\eqref{e:MOND}, with the boundary conditions given by the constant galactic gravitational field $\bm{g}_\text{e}$ (see Fig.~\ref{f:setting}), i.e. $\lim_{r\to \infty} \bm{g} = \bm{g}_\text{e} = g_\text{e} \bm{e}$. In order to be closer to a Poisson-like form of the partial differential equation (PDE), we are solving in terms of the internal potential $u = U - \bm{g}_\text{e} \cdot \mathbf{x}$ such that $\lim_{r\to \infty}u = 0$. Defining $\bm{h} = \vecnab u = \bm{g} - \bm{g}_\text{e}$, the MOND equation~\eqref{e:MOND} becomes the following PDE:
\begin{equation}
  \label{e:MOND_LORENE}
  \Delta u = \frac{1}{\mu(g/a_0)} \left\{ -4\pi G
    \rho - \frac{\mu'(g/a_0)}{a_0 g} \,g^i g^j \partial_i
   h_j \right\}\,, 
\end{equation}
with $\mu' \equiv \ud \mu / \ud y$ being the derivative of the function $\mu$ with respect to its argument $y=g/a_0$. 

This PDE is solved numerically, using the library \textsc{lorene} which implements spectral methods (for a review in the case of numerical relativity see \citealt{GN09}) in spherical coordinates. In our case (see Fig.~\ref{f:setting}), the fields do not depend on the azimuthal angle $\varphi$ and the problem is axisymmetric. Since Eq.~\eqref{e:MOND_LORENE} is a non-linear elliptic PDE, the algorithm used is the fixed point iteration method in which one starts from an initial guess $u_0(r,\th)$, here the solution of the Newtonian Poisson equation $\Delta u_0 = -4\pi G \rho$. Knowing $u_n(r, \th)$ at a given iteration step $n$, one computes the non-linear source terms in the right-hand-side of Eq.~\eqref{e:MOND_LORENE}, say $\sigma(u,\rho)$, to obtain a new value of the potential $u_{n+1}$, solving a linear Poisson equation
\begin{equation}
  \label{e:linear_MOND}
  \Delta \tilde{u}_{n+1} = \sigma(u_n, \rho)\,,
\end{equation}
and using a relaxation technique
\begin{equation}
  \label{e:relax_iteration}
  u_{n+1} = \lambda \tilde{u}_{n+1} + (1 - \lambda) u_n\,,
\end{equation}
with $\lambda \in (0,1]$ being the relaxation parameter (often taken to be $0.5$). This iteration is stopped when the relative variation of the potential $u$ between two successive steps becomes lower than a given threshold (usually $10^{-12}$).

The linear equation~\eqref{e:linear_MOND} is solved decomposing each field on a truncated base of spherical harmonics $Y^m_l(\th,\varphi)$ (with $m=0$ because the problem is axisymmetric) in the $\th$-direction, and Chebyshev polynomials $T_n(r)$ in the $r$-direction. For this last coordinate, a multi-domain technique is used, with linear mappings of the $r$ coordinate to the interval $[-1,1]$, except for the last domain where a mapping of the type $s=1/r$ allows to treat spatial infinity in our computational domain and to impose the right boundary conditions at $r\to \infty$ (see \citealt{GN09} for details). Moreover, this multi-domain technique imposes that the potential $u$, as well as its first radial derivative, be continuous across the stellar surface.

Finally, the only modification made to the exact equation~\eqref{e:MOND_LORENE} in order to be numerically integrated is the setting $\mu(g/a_0)=1$ inside the star. Indeed, there are two problems here. First, the MOND gravitational field $g(r)$ does not admit a regular Taylor expansion near the origin, as does the density in Eq.~\eqref{e:Taylor_rho_spher} and the associated Newtonian gravitational field which follows as
\begin{equation}
g_\text{N} = 4\pi G \sum_{k=0}^{+\infty}\frac{r^{2k+1}}{(2k+3)(2k+1)!} \,(\Delta^k\rho)(\mathbf{0})\,.
\label{e:Taylor_g_spher}
\end{equation}
This can be immediately seen from the fact that in spherical symmetry the MOND equation \eqref{e:MOND} reduces to $\mu(g/a_0)g=g_\text{N}$. Using \eqref{e:Taylor_g_spher} we see that the expansion of $g$ when $r\to 0$ starts with a term proportional to the square root of $r$ and is therefore not regular; more precisely we have
\begin{equation}
g = \sqrt{\frac{4\pi G\rho_0a_0r}{3}}\Bigl[1+\mathcal{O}(r^2)\Bigr]\,.
\end{equation}
With our polynomial decomposition of fields in the radial direction, where we decompose in Chebyshev polynomials $T_n(r)$, we therefore get an incompatibility when trying to solve the Poisson equation~\eqref{e:linear_MOND} in the vicinity of the origin at the centre of the Sun. The second problem is that, at the very centre of the star $g \to 0$ and therefore $\mu \to 0$, which makes the division present in the right-hand-side of Eq.~\eqref{e:MOND_LORENE} difficult to handle numerically. 

Here we have chosen to avoid both problems by imposing $\mu=1$ in the star so it is entirely Newtonian. This approximation may produce a noticeable change in the value of $\mu(y)$ only within a sphere of radius $r \lesssim \frac{3a_0}{4\pi G \rho_0} \sim 10^{-15} R_\odot$, where $\rho_0$ is the central density and $R_\odot$ is the radius of the Sun. It is thus supposed to have a completely negligible effect on the results.

The result which is finally used to study the influence of the modification of the Newtonian gravity on the orbits of planets is the value of the trace-free multipole moments $Q_L$ defined in Sec.~\ref{s:multipoles}. In the case where all the multipole moments are induced by the presence of the external field $\bm{g}_\text{e}$ in the preferred direction $\bm{e}$, the situation is axisymmetric and all the moments $Q_L$ will have their axis pointing in that direction $\bm{e}$, and we can define the multipole coefficients $Q_l$ by
\begin{equation}
  \label{QLeL}
Q_L = Q_l \,\hat{e}^L\,,
\end{equation}
where $\hat{e}^L$ denotes the STF part of the product of $l$ unit vectors $e^L=e^{i_1}\cdots e^{i_l}$. Then the multipole expansion \eqref{multexp} reads as\footnote{Here, the expansion is defined for any range in radius $r$, contrary to Sec.~\ref{s:multipoles} and Appendix~\ref{appB}, where the multipoles were only numbers defined for $r\to 0$ and not functions of $r$.}
\begin{equation}
  \label{legendre}
\delta u(r, \th) = \sum_{l=0}^{+\infty} \frac{(-)^l}{(2l-1)!!} \,r^l \,Q_l(r) \,P_l (\cos\th)\,,
\end{equation}
where $P_l$ is the usual Legendre polynomial and $\th$ is defined in Fig.~\ref{f:setting}. Although from the considerations of Sec.~\ref{s:multipoles} the multipole moments $Q_l$ should be approximately constant within the MOND transition radius $r_0$, here we compute them directly from the numerical solution of \eqref{e:MOND} and shall obtain their dependence on $r$; we shall check numerically that $Q_2(r)$ and $Q_3(r)$ are indeed almost constant in a large sphere surrounding the Solar system. With our definition the monopole, quadrupole and octupole pieces in the internal field are given by
\begin{subequations}\label{e:def_multipole}
\begin{align}
  \label{e:def_dipole}
\delta u_1 &= - r \,Q_1 \,\cos\th \,,\\
\label{e:def_quadrupole}\delta u_2 &= \frac{1}{2} \,r^2 \,Q_2 \left(\cos^2\th - \frac{1}{3}\right)\,,\\
\label{e:def_octupole}\delta u_3 &= -\frac{1}{6} \,r^3 \,Q_3 \,\cos\th\left(\cos^3\th - \frac{3}{5} \cos\th\right)\,.
\end{align}\end{subequations}
From dimensional arguments, it is possible to see that $Q_l$ should scale as $Q_l \sim a_0/r_0^{l-1}$ [see for instance Eqs.~\eqref{q1} and \eqref{q23}] and therefore\footnote{The radial dependence of the contribution of the $l$-th multipole moment is the same as that of the usual MOND spherical effect corresponding to a MOND function behaving like $\mu\sim 1-\epsilon y^{-\gamma}$ when $y\to\infty$, with $\gamma=(l+1)/2$ (see the discussion at the end of Sec.~\ref{ss:motiv}). But of course the effect we are considering here is non-spherical.}
\begin{equation}
  \label{e:eq_delta_u}
  \delta u \sim \sum_{l} Q_l r^l \sim a_0 r_0 \sum_{l} \left( \frac{r}{r_0}\right)^l .
\end{equation}
This last expression shows that higher-order multipoles should have lower influence on Solar-system planets, for which $r\ll r_0$. This shall be exemplified with the influence of the octupole, as compared to that of the quadrupole, in Sec.~\ref{ss:precession}.

\subsection{Tests of the code}\label{ss:tests}

A first test of our code is the standard comparison between the Laplace operator applied to the potential $u$, and the right-hand-side of Eq.~\eqref{e:MOND_LORENE}. In all results displayed here, the maximum of this error has always remained lower than $10^{-11}$ and shall not be considered as an interesting error indicator. Next, an indication of the error comes from the use of the Gauss theorem:
\begin{equation}
  \label{e:Gauss}
  \oint \mu\left( \frac{g}{a_0} \right) \bm{g} \cdot \textrm{d}^2\bm{S} = -4\pi G M, 
\end{equation}
where $M$ is the mass of the star, computed from initial data at a very high precision. Then, another check (which is however not \textit{a priori} independent) is the asymptotic behaviour of the potential $u(r,\th)$ when $r \to \infty$, which is 
\begin{equation}
  \label{e:asym_check}
  u = \frac{GM}{r\mu_\text{e}\sqrt{1 + \lambda_\text{e} \sin^2 \th}} + \mathcal{O} \left(
    \frac{1}{r^2} \right)\,,
\end{equation}
where $\mu_\text{e} = \mu(y_\text{e})$ and $\lambda_\text{e} = y_\text{e} \mu'_\text{e}/\mu_\text{e}$ with $y_\text{e}=g_\text{e}/a_0$ (see the proof in Appendix~\ref{appA}). The two tests are not really independent because the Gauss theorem is obtained by integrating the asymptotic field over a sphere at infinity. The main interest of these tests is that while the field is computed asymptotically on a sphere at infinity, the mass $M$ is obtained ``locally'' by integrating numerically the density over the volume of the star. We emphasise that in our code the correct asymptotic behaviour \eqref{e:asym_check} comes out directly and needs not to be imposed by hands at the beginning of the calculation.
\begin{figure}
  \centerline{\includegraphics[height=9cm,angle=-90]{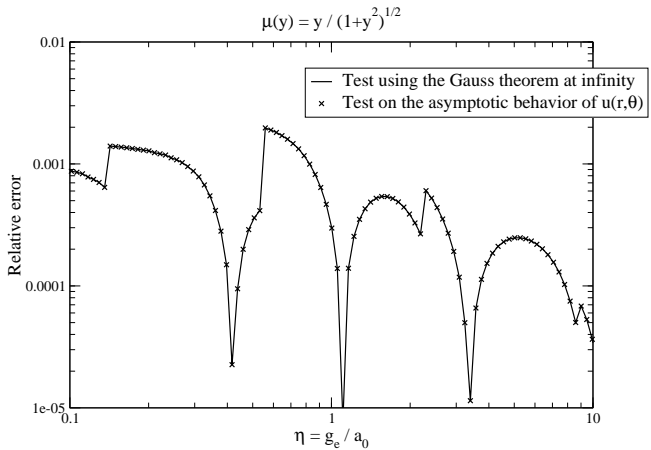}}\vspace{0.5em}

  \caption{Relative error of the code on the two tests described in
    Sec.~\ref{ss:tests}, for a $1 M_\odot$ spherical mass distribution,
    for the standard MOND function $\mu_2(y)$ [Eq.~\eqref{mu2}] and the standard value $a_0 = 1.2 \times 10^{-10} \textrm{m.s}^{-2}$. The abscissa gives the ratio \eqref{eta} between the norm of the asymptotic gravitational field $g_\text{e}$ and the parameter $a_0$. Note that both curves do coincide.}
  \label{f:precision}
\end{figure}

From the use of the mapping $s=1/r$ (see Sec.~\ref{ss:implementation}), it is numerically possible to check the angular dependence of the quantity $r\times u$ when $r\to \infty$ and thus to estimate independently the error on the potential $u$. Both tests --- Gauss theorem~\eqref{e:Gauss} and asymptotic behaviour~\eqref{e:asym_check} --- have been performed and have given accuracy levels of $\sim 10^{-3}$--$10^{-4}$. Some examples are given in Fig.~\ref{f:precision}, for a sequence of different values for the asymptotic gravitational field $g_\text{e}$. Both tests give the same numbers, up to five digits. In all results shown hereafter, the accuracy level defined by these two tests is better than $10^{-2}$.

\begin{figure}
  {}\vspace{2em}
  \centerline{\includegraphics[height=9cm,angle=-90]{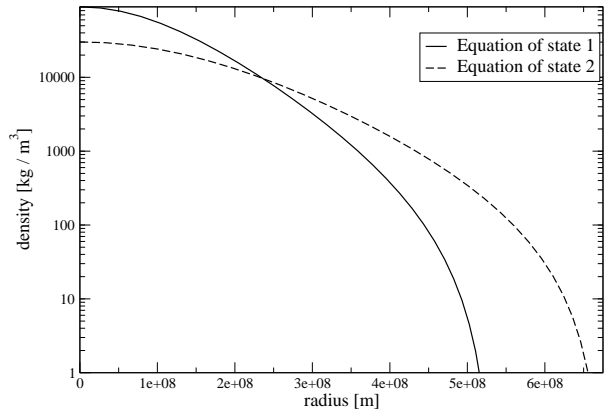}}\vspace{0.5em}
  \caption{Two density profiles for testing the result dependence on the
    equation of state. Both profiles yield a one-solar mass star in
    hydrostatic equilibrium. The two profiles yield very little (same level as
    the overall numerical accuracy $\sim 10^{-3}$) difference on our results.}
  \label{f:eos}
\end{figure}
Among the parameters entering the numerical model, we have checked that the details of the density profile $\rho(r)$ do not contribute to the value of the quadrupole moment $Q_2(r)$. As an illustration, we give here the quadrupole obtained with two different density profiles, displayed in Fig.~\ref{f:eos}, both giving a star of exactly one solar mass. With the same other parameters [choices of $\mu(y)$, $a_0$ and $g_\text{e}$], the relative difference in the induced quadrupole is $4\times 10^{-3}$, while the error indicators give a relative numerical uncertainty of $\sim 2\times 10^{-3}$. We therefore consider that the particular form of the density profile used to model the star (i.e., the choice of EOS) has little influence on the results.

\subsection{Results on the induced quadrupole and octupole}
\label{ss:quadrupole}

\begin{figure*}
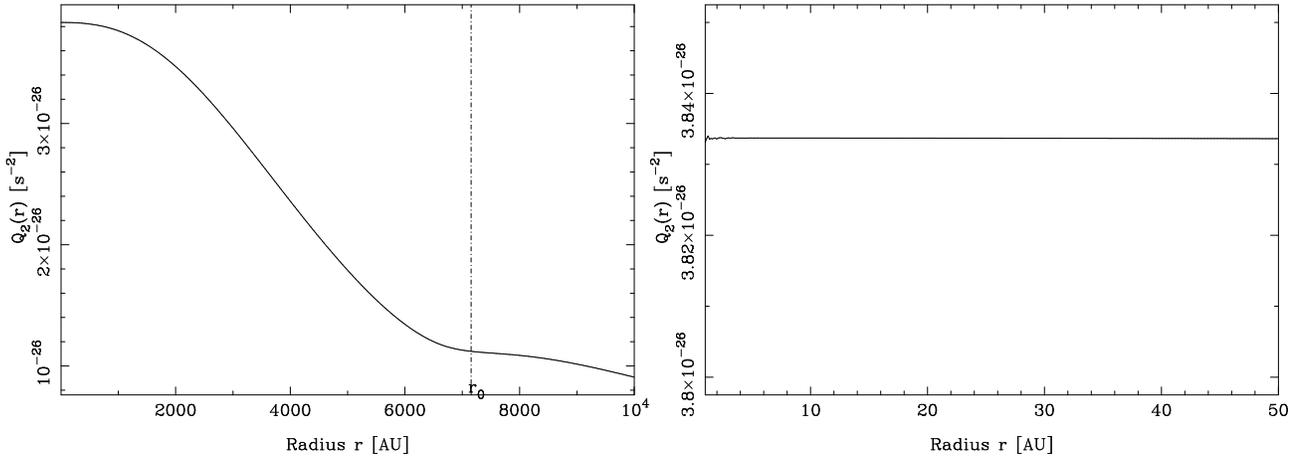

  \centerline{\includegraphics[width=6cm,angle=-90]{q_r_alpha1.ps}\includegraphics[width=6cm,angle=-90]{qr_zoom.ps}}
  \caption{Left panel: profile of $Q_2(r)$ in the solar system, for a standard choice of function $\mu_1(y)$ [see Eq.~\eqref{mu1}], $a_0 = 1.2 \times 10^{-10} \textrm{m.s}^{-2}$ and $g_\text{e} = 1.9 \times 10^{-10} \textrm{ m.s}^{-2}$. The MOND transition radius $r_0 = \sqrt{GM/a_0}$ is shown by a dash-dotted line at $r\simeq 7100$ AU. Right panel: zoom of the central region ($r\leqslant 50$ AU), where the quadrupole is almost constant (the spikes for $r\to 0$ are due to the high value of the monopole term, from which it is numerically difficult to extract the quadrupole). Note the difference in the y-axis scales.}
  \label{f:qralpha1}
\end{figure*}
In what follows, unless specified, the mass of the star is that of the Sun. As a first result, we show in Fig.~\ref{f:qralpha1} the profile of the quadrupole induced by the MOND theory through the function $Q_2(r)$ defined in Eq.~\eqref{e:def_quadrupole}. We find that this quadrupole is decreasing from the star's neighbourhood to zero, on a typical scale of $10000$ astronomical units (AU). However, this quadrupole can be considered as almost constant closer to the Sun, as it has a relative variation lower than $10^{-4}$ within 30 AU (see the zoomed region in Fig.~\ref{f:qralpha1}). We shall therefore refer to the quadrupole as a simple number, noted $Q_2(0)$ or simply $Q_2$, when evaluating its influence on the orbits of Solar-system planets. Our numerical results for the quadrupole are given in Table~\ref{tab1}. 

We now briefly study other multipole moments, namely for $l=1$ and $l=3$.
\begin{figure*}
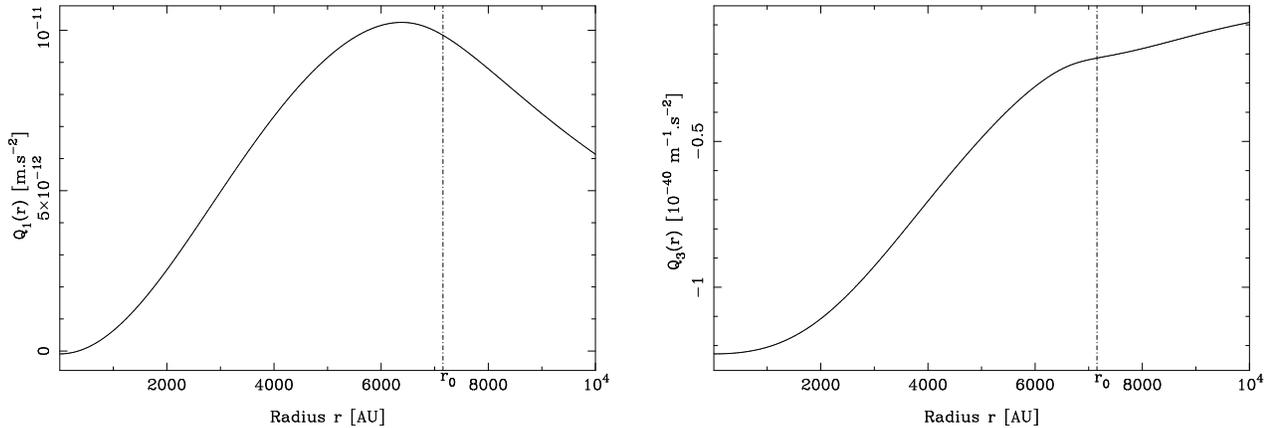

  \centerline{\includegraphics[height=8cm,angle=-90]{q1r_alpha1.ps}\hspace{2em}
  \includegraphics[height=8cm,angle=-90]{q3r_alpha1.ps}}
\caption{Profiles of dipole $Q_1(r)$ (left) and octupole $Q_3(r)$ (right) in the solar system, with same settings as those of Fig.~\ref{f:qralpha1}. The MOND transition radius $r_0 = \sqrt{GM/a_0}$ is shown by a dash-dotted line at $r\simeq 7100$ AU. Numerically we get $\left|Q_1(0)\right| \sim 9 \times 10^{-14} \,\textrm{m.s}^{-2} \ll a_0$ which shows that the dipole is in fact zero.}
\label{f:qr_l1l3}
\end{figure*}
The profiles for the dipole $Q_1(r)$ and octupole $Q_3(r)$, defined by~\eqref{e:def_dipole} and~\eqref{e:def_octupole}, are displayed in Fig.~\ref{f:qr_l1l3}. We find that near the Sun and the solar planets the dipole can be considered as zero, since it is three orders of magnitude lower than the typical value for an acceleration in the problem (i.e. $a_0$ or $g_\text{e}$). Indeed we expect on dimensional analysis that $Q_1$ should scale with the MOND acceleration $a_0$ [see~\eqref{e:eq_delta_u}],
\begin{equation}\label{q1}
Q_1 = a_0\,q_1(\eta)\,,
\end{equation}
where the dimensionless coefficient $q_1$ depends on the ratio $\eta$ between $g_\text{e}$ and $a_0$ as defined by Eq.~\eqref{eta}. Our values given in Table~\ref{tab1} show that $q_1$ is actually very small. It means that $Q_1$ is lower than the numerical error and confirms the analytical proof given in Sec.~\ref{s:multipoles} that the dipole moment is approximately zero. On the other hand, we find that the octupole tends to a non-zero value, because the corresponding $q_3$, which has also an influence on the orbits of planets (see the next section), is found to be of the order of one. Numerical values of the octupole are also given in Table~\ref{tab1}.
\begin{figure}
  \centerline{\includegraphics[height= 9cm,angle=-90]{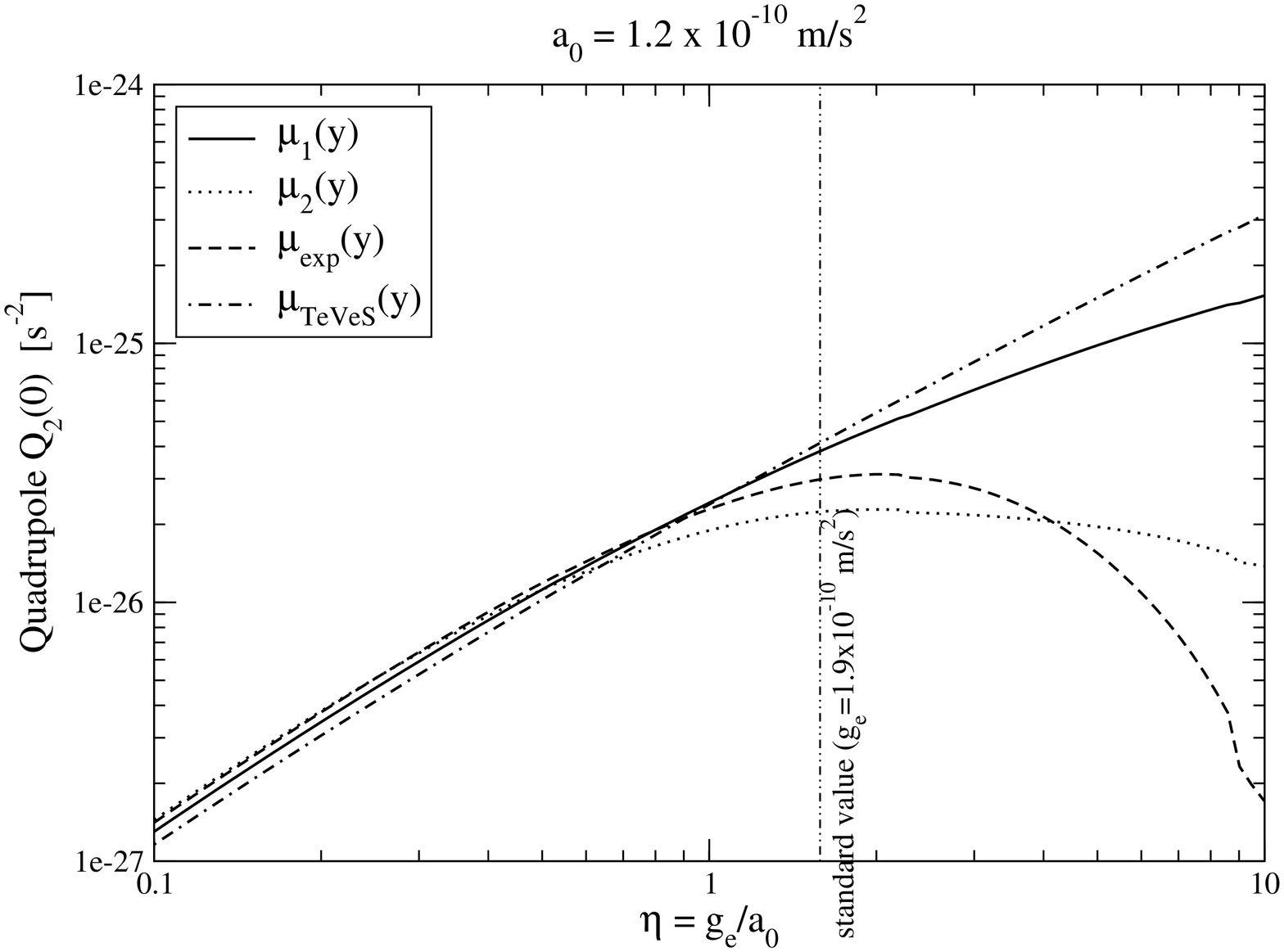}}\vspace{0.5em}
  \caption{Quadrupole in the solar system $Q_2\equiv Q_2(0)$ as a function of the external galactic gravitational field $g_\text{e}$, for four different expressions of the coupling function $\mu(y)$ [see Eqs.~\eqref{mu} for a description]. The standard value of $g_\text{e} = 1.9\times 10^{-10} \textrm{ m.s}^{-2}$ is shown by a thin dash-double-dotted line.}
  \label{f:qge}
\end{figure}

The dependence of the quadrupole moment $Q_2$ upon the value of the galactic gravitational field is displayed in Fig.~\ref{f:qge}, for four different coupling functions $\mu(y)$. Here we consider four cases widely used in the literature:
\begin{subequations}\label{mu}
\begin{align}
\mu_1(y) &= \dfrac{y}{1+y}\,,\label{mu1}\\
\mu_2(y) &= \dfrac{y}{\sqrt{1+y^2}}\,,\label{mu2}\\
\mu_{\textrm{exp}}(y) &= 1-e^{-y}\,,\label{mu3}\\
\mu_{\textrm{TeVeS}}(y) &= \dfrac{\sqrt{1+4y} - 1}{\sqrt{1+4y} + 1}\,.\label{mu4}
\end{align}
\end{subequations}
The function $\mu_1$ has been shown to yield good fits of galactic rotation curves \citep{FB05}. However because of its slow transition to the Newtonian regime it is \textit{a priori} incompatible with Solar System observations. The function $\mu_2$ is generally called the ``standard'' choice and was used in fits such as those of \cite{BBS91}. We here use the general notation for any positive integer $n$:
\begin{equation}
  \label{e:def_mu_n}
  \mu_n(y) = \frac{y}{\sqrt[n]{1+y^n}}.
\end{equation}
We include also the function $\mu_{\textrm{exp}}$ having an exponentially fast transition to the Newtonian regime. The fourth choice $\mu_{\textrm{TeVeS}}$ is motivated by the theory TeVeS \citep{Bek04}. One should note that none of these functions, except maybe the fourth one, derives from a fundamental physical principle.

One observes that, up to the standard value of $g_\text{e} = 1.9\times 10^{-10} \textrm{ m.s}^{-2}$ the four curves are quite close, giving an interval of values for $Q_2$:
\begin{equation}
  \label{e:interv_Q}
  2.2\times 10^{-26} \textrm{ s}^{-2} \lesssim Q_2 \lesssim 4.1\times
10^{-26} \textrm{ s}^{-2}.
\end{equation}
However, for other choices of the MOND function $\mu(y)$, $Q_2$ can take lower values down to $2.1\times 10^{-27} \textrm{ s}^{-2}$. In all cases, we find that $Q_2$ is positive, which means a prolate deformation of the field toward the galactic centre.
Two profiles of $\mu(y)$ seem to give a maximum for $Q_2$ near the standard value of $g_\text{e}$, namely the standard choice $\mu_2$ and the exponential choice $\mu_{\textrm{exp}}$. For the two other choices of $\mu(y)$, we have not been able to increase the value of $\eta = g_\text{e} / a_0$ further than $10$, because the error indicators would become too large and the results could not be trusted any longer. Note that in Fig.~\ref{f:qge}, $a_0$ has been kept fixed, while we have varied $g_\text{e}$. We have further explored the dependence of the EFE induced quadrupole $Q_2$ on the type of MOND function. 

\begin{figure}
  {}\vspace{1em}
  \centerline{\includegraphics[height=9cm,angle=-90]{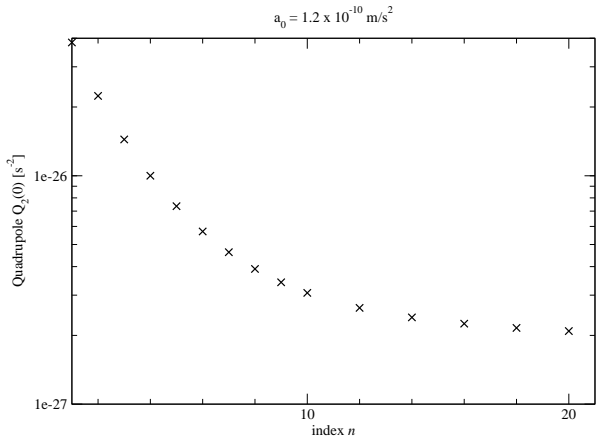}}\vspace{0.5em}
  \caption{Quadrupole moment $Q_2$ as a function of the index $n$ of the MOND function $\mu_n(y)$ defined by Eq.~(\ref{e:def_mu_n}). \label{f:q2_mun}}
\end{figure}

We have used several different functions of type $\mu_n$, as defined in Eq.~(\ref{e:def_mu_n}), and the results for the variation of $Q_2$ depending on $n$ are displayed in Fig.~\ref{f:q2_mun}. From this figure, one can notice that the value of $Q_2$ decreases with $n$, that is with a faster transition from the weak-field regime (where $\mu(y) \sim y$) to the strong field regime ($\mu(y) \sim 1$). We have been unable to study higher values of $n$ and to determine a possible limit for $Q_2$ as $n$ goes to infinity.
In Table~\ref{tab1}, we give values of the dipole, quadrupole and octupole near the Sun ($r\leqslant 50$ AU), where they are observed to be constant, for the four different expressions \eqref{mu} of the interpolating function. 
\begin{table*}
  \centering
  \caption{Numerical values of the dipole $Q_1$, the quadrupole $Q_2$ and the octupole $Q_3$ together with the associated dimensionless quantities $q_1, q_2$ and $q_3$ defined by Eqs.~\eqref{q1} and~\eqref{q23}. All values are given near the Sun. We use different choices of the function $\mu(y)$ defined in Eqs.~\eqref{mu} and~\eqref{e:def_mu_n}. These values have been obtained for $a_0 = 1.2\times 10^{-10} \textrm{m}\cdot\text{s}^{-2}$ and $g_\text{e} = 1.9\times 10^{-10} \textrm{m}\cdot\text{s}^{-2}$ (and $M=1\,M_\odot$).}
    \label{tab1}
  \begin{tabular}{lcccccc}
\hline
MOND function & $\mu_1(y)$ & $\mu_2(y)$ & $\mu_5(y)$ & $\mu_{20}(y)$ & $\mu_{\textrm{exp}}(y)$ & 
    $\mu_{\textrm{TeVeS}}(y)$\\
    \hline
    $Q_1$ [$\text{m}\,\text{s}^{-2}$] & $-8.9\times 10^{-14}$ & $-9.8\times 10^{-14}$ & $-1.1\times 10^{-13}$ & $-1.2\times 10^{-13}$ &$-9.7\times 10^{-14}$ &$-3.5\times 10^{-14}$ \\
    $q_1$ & $-7.4\times 10^{-4}$ & $-8.2 \times 10^{-4}$ & $-9.2\times 10^{-4}$ & $-10^{-3}$ & $-8.1 \times 10^{-4}$ &$-2.9\times 10^{-4}$\\
    \hline
    $Q_2$ [$\text{s}^{-2}$] & $3.8\times 10^{-26}$ & $2.2\times 10^{-26}$ & $7.4\times 10^{-27}$ & $2.1\times10^{-27}$& $3.0\times 10^{-26}$ &$4.1\times 10^{-26}$ \\
    $q_2$ & $0.33$ &$0.19$ &$6.5\times10^{-2}$ & $1.8\times 10^{-2}$ & $0.26$ &$0.36$\\
    \hline
    $Q_3$ [$\text{m}^{-1}\text{s}^{-2}$] & $-1.2\times 10^{-40}$ &$-9.3\times 10^{-41}$ & $-4.9\times 10^{-42}$ & $-2.3 \times 10^{-42}$ & $-1.2\times 10^{-40}$ &$-1.1\times 10^{-40}$ \\
    $q_3$ & $-1.1$ &$-0.87$ & $-4.6\times 10^{-2}$ & $-2.1\times 10^{-2}$ & $-1.1$ &$-1$ \\
\hline
  \end{tabular}
\end{table*}
\begin{figure}
  {}\vspace{1em}
  \centerline{\includegraphics[height=9cm,angle=-90]{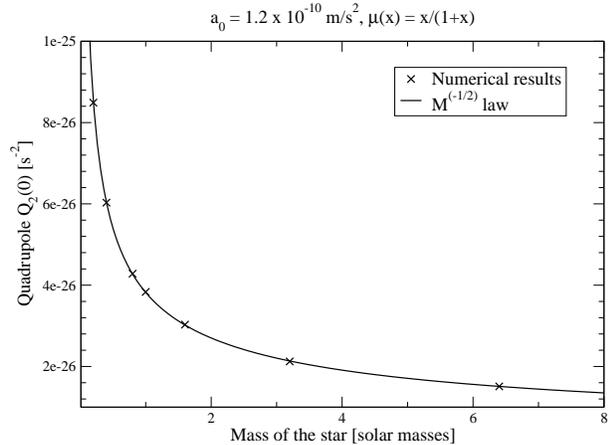}}\vspace{0.5em}
  \caption{Quadrupole moment $Q_2$ as a function of the stellar mass $M$. The $M^{-1/2}$-law is recovered as expected from Eq.~\eqref{q2}. \label{f:qmass}}
\end{figure}

We have also explored different values of $a_0$ at fixed $g_\text{e}$, with the results that $Q_2$ was increasing for small $a_0$ as a power-law, reaching a maximum value for $a_0 \simeq 10^{-10} - 10^{-9} \textrm{m}\cdot\textrm{s}^{-2}$ and then slowly decreasing. Finally, we have varied the mass of the central star. Results are displayed in Fig.~\ref{f:qmass} where, in particular, the $M^{-1/2}$ dependence of the quadrupole moment is recovered. This is not a surprise since on general grounds one expects that the quadrupole moment (and similarly the octupole moment) should scale like \citep{Milg09}\footnote{To compare with the results of \cite{Milg09} for the quadrupole, one should note the different conventions in use, namely $q^\text{Milgrom}_{ij}=-Q^\text{BN}_{ij}$ and $q^\text{Milgrom}(\eta)=-\frac{2}{3}q^\text{BN}_2(\eta)$. Our results for the quadrupole are in good agreement with values given by \cite{Milg09}.}
\begin{subequations}\label{q23}
\begin{align}
Q_2 &= \frac{a_0}{r_0} \,q_2(\eta)\,,\label{q2}\\
Q_3 &= \frac{a_0}{r_0^2} \,q_3(\eta)\,,\label{q3}
\end{align}\end{subequations}
where $r_0=\sqrt{G M/a_0}$ is the MOND transition radius and $q_2$, $q_3$ denote some dimensionless coefficients depending on the ratio $\eta=g_\text{e}/a_0$ and on the choice of the interpolating function $\mu$. Having obtained the expected dependence on the mass in Fig.~\ref{f:qmass} is another check that our code behaves correctly, since we are able to recover the analytic results known for this system. The values of the dimensionless multipole coefficients $q_2$ and $q_3$, which are expectedly close to unity, are reported in Table~\ref{tab1}.

\section{Effect on the dynamics of solar system planets}\label{s:effect}

\subsection{Perturbation equations}\label{ss:perturbation}

In this section, we investigate the consequence for the dynamics of inner planets of the Solar System of the presence of an abnormal multipole moment $Q_L$ oriented toward the direction $\bm{e}$ of the galactic centre, in the sense of Eq.~\eqref{QLeL}. Recall that the domain of validity of this anomaly is expected to enclose all the inner Solar System (for distances $r\lesssim r_0\approx 7100$ AU), with the quadrupole coefficient being constant up to say $50$ AU (see Fig.~\ref{f:qralpha1}). As we have seen, the anomaly induces a perturbation on the Newtonian gravitational potential, namely $u=u_\text{N}+\delta u$, where the perturbation function $\delta u$ is given for various multipole moments by Eqs.~\eqref{e:def_multipole}. Following the standard practice of celestial mechanics we denote the perturbation by $R\equiv\delta u$. The perturbation function $R$ is such that the perturbing force (or, rather, acceleration) acting on the Newtonian motion is $\bm{F} = \vecnab R$.

The unperturbed Keplerian orbit of a planet around the Sun is described by six orbital elements (say $c_A$ with $A=1,\cdots,6$). For these we adopt the semi-major axis $a$, the eccentricity $e$, the inclination $I$ of the orbital plane, the mean anomaly $\ell$ defined by $\ell=n(t-T)$ where $n=2\pi/P$ ($n$ is the mean motion, $P$ is the orbital period and $T$ is the instant of passage at the perihelion), the argument of the perihelion $\omega$ (or angular distance from ascending node to perihelion), and the longitude of the ascending node $\Omega$. We also use the longitude of the perihelion defined by $\tilde{\omega}=\omega+\Omega$.

The perturbation is a function $R(c_A)$ of the orbital elements of the unperturbed Keplerian ellipse. With our choice for the orbital elements $\{c_A\}=\{a,e,I,\ell,\omega,\Omega\}$ the perturbation equations of celestial mechanics read as (see e.g. \citealt{BrouwerClemence})
\begin{subequations}\label{perteqs}
\begin{eqnarray}
\frac{\ud a}{\ud t} &=& \frac{2}{a n} \frac{\partial R}{\partial \ell}\,,\label{perteqa}\\
\frac{\ud e}{\ud t} &=& \frac{\sqrt{1-e^2}}{e a^2 n} \left[ \sqrt{1-e^2} \frac{\partial R}{\partial \ell} - \frac{\partial R}{\partial \omega} \right]\,,\\
\frac{\ud I}{\ud t} &=& \frac{1}{a^2 n \sqrt{1-e^2} \sin I} \left[ \cos I \frac{\partial R}{\partial \omega} - \frac{\partial R}{\partial \Omega} \right]\,,\\
\frac{\ud \ell}{\ud t} &=& n - \frac{1}{a^2 n} \left[ 2 a \frac{\partial R}{\partial a} + \frac{1-e^2}{e} \frac{\partial R}{\partial e} \right]\,,\\
\frac{\ud \omega}{\ud t} &=& \frac{1}{a^2 n} \left[ \frac{\sqrt{1-e^2}}{e} \frac{\partial R}{\partial e} - \frac{\cos I}{\sin I \sqrt{1-e^2}} \frac{\partial R}{\partial I} \right]\,,\\
\frac{\ud \Omega}{\ud t} &=& \frac{1}{a^2 n \sin I \sqrt{1-e^2}} \frac{\partial R}{\partial I}\,.
\end{eqnarray}\end{subequations}

We shall be interested only in secular effects, so we average these equations over one orbital period $P$; denoting the time average by brackets we have
\begin{equation}\label{av}
\left\langle\frac{\partial R}{\partial c_A}\right\rangle = \frac{1}{P} \int_0^P \ud t \,\frac{\partial R}{\partial c_A} = \frac{1}{2\pi}\int_0^{2\pi} \ud \ell \,\frac{\partial R}{\partial c_A}\,.
\end{equation}
In particular this implies that we shall always have $\langle\partial R/\partial \ell\rangle =0$ hence the perturbation equation~\eqref{perteqa} gives $\langle\ud a/\ud t\rangle =0$ (at first order in perturbation theory). Indeed, a perturbative force of the type $\bm{F} = \vecnab R$ is conservative, so the energy of the orbit is conserved in average and there is no secular change in the semi-major axis.

Let us now apply the perturbation equations \eqref{perteqs} to the specific case of the perturbation function corresponding to the quadrupole anomaly, namely
\begin{equation}\label{R2}
R = \delta u_2 = \frac{1}{2} \,r^2 \,Q_2 \left(\cos^2\th - \frac{1}{3}\right)\,.
\end{equation}
Here $\cos\th = \bm{e}\cdot\mathbf{n}$, with $\mathbf{n}=\mathbf{x}/r$ being the unit direction of the planet and $r^2=x^2+y^2+z^2$, where $(x,y,z)$ are the coordinates of the planet in an absolute Galilean frame centred on the focus of the unperturbed Keplerian ellipse, and with respect to which the orbital elements $\{a,e,I,\ell,\omega,\Omega\}$ are defined. In the following, to simplify the presentation, we shall choose the $x$-direction of this Galilean frame to be the direction of the galactic centre $\bm{e}=\bm{g}_\text{e}/g_\text{e}$. That is, we assume that the origin of the longitude of the ascending node $\Omega$ lies in the direction of the galactic centre. This means that $\cos\th=x/r$ so that
\begin{equation}\label{R2'}
R = \frac{Q_2}{6} \Bigl(2x^2-y^2-z^2\Bigr)\,.
\end{equation}
We express the planet's absolute coordinates $(x,y,z)$ in terms of the orbital elements $\{a,e,I,\ell,\omega,\Omega\}$ by performing as usual three successive frame rotations with angles $\Omega$, $I$ and $\omega$, to arrive at the frame $(u,v,w)$ associated with the motion, where $(u,v)$ is in the orbital plane, with $u$ in the direction of the perihelion and $v$ oriented in the sense of motion at perihelion. The unperturbed coordinates of the planet in this frame are
\begin{subequations}\label{comp}
\begin{eqnarray}
u &=& a \left(\cos U -e\right)\,,\\
v &=& a \sqrt{1-e^2} \sin U\,,\\
w &=& 0\,,
\end{eqnarray}\end{subequations}
where $U$ denotes the eccentric anomaly, related to $\ell$ by the Kepler equation $\ell = U - e \sin U$. Inserting the resulting expression of $R$ into the perturbation equations \eqref{perteqs}, and performing the time average (in practice, an average over the eccentric anomaly $U$, after the appropriate change of variable), then yields
\begin{subequations}\label{avperteqs}
\begin{eqnarray}
\left\langle\frac{\ud a}{\ud t}\right\rangle &=& 0\,,\\
\left\langle\frac{\ud e}{\ud t}\right\rangle &=& \frac{5 Q_2 e \sqrt{1-e^2}}{4 n} \left[ \cos I \cos(2\omega)\sin(2\Omega)\right.\nonumber\\
  &&\qquad\left. +\sin(2\omega)(\cos^2\Omega-\cos^2I\sin^2\Omega)\right]\,,\\
\left\langle\frac{\ud I}{\ud t}\right\rangle &=& \frac{Q_2 \sin I\sin\Omega}{4 n \sqrt{1-e^2}} \left[ \left(2+3e^2+5e^2\cos(2\omega)\right)\cos\Omega\right.\nonumber \\
&&\qquad\left. - 5 e^2\cos I\sin(2\omega)\sin\Omega\right]\,,\\
\left\langle\frac{\ud \ell}{\ud t}\right\rangle &=& n + \frac{Q_2}{96 n} \left\{ -15(1+e^2)\cos(2\omega)\left[2-2\cos(2I)\right.\right.\nonumber\\&&\left.\qquad+\cos(2I-2\Omega)+6\cos(2\Omega)+\cos(2I+2\Omega)\right]\nonumber\\&&+12\left(-8+3e^2\right)\cos^2\Omega\nonumber\\&&-12\left(1-6e^2+(7+3e^2)\cos(2I)\right)\sin^2\Omega\nonumber\\&&\left.+20\left(2-3e^2\right.\right.\nonumber\\
&&\qquad\left.\left.+6(1+e^2)\cos I\sin(2\omega)\sin(2\Omega)\right)\right\}\,,\\
\left\langle\frac{\ud \omega}{\ud t}\right\rangle &=& -\frac{Q_2}{4n \sqrt{1-e^2}} \left[ 2 - 2 e^2\right.\nonumber\\&&\qquad\left.+(-1+e^2)\left(3+5\cos(2\omega)\right)\cos^2\Omega\right.\nonumber\\&&\qquad\left.+\frac{5}{2}e^2\cos I\sin(2\omega)\sin(2\Omega)\right.\nonumber\\&&\qquad\left.-10\cos^2I\sin^2\omega\sin^2\Omega\right.\nonumber\\&&\qquad\left.+5(1-e^2)\cos I\sin(2\omega)\sin(2\Omega)\right]\,,\\
\left\langle\frac{\ud \Omega}{\ud t}\right\rangle &=& \frac{Q_2\sin\Omega}{4n \sqrt{1-e^2}}\left[ 5e^2\cos\Omega\sin(2\omega)\right.\nonumber\\
&&\qquad\left.+ \cos I\left(-2-3e^2+5e^2\cos(2\omega)\right)\sin\Omega\right]\,.
\end{eqnarray}\end{subequations}
These equations are general and give in particular the precession of the node $\langle\ud \Omega/\ud t\rangle$ and the perihelion precession $\langle\ud \omega/\ud t\rangle$ or, rather, $\langle\ud \tilde{\omega}/\ud t\rangle$ where $\tilde{\omega}=\omega+\Omega$ is the longitude of the perihelion.

In order to make some estimate of the magnitude of the effect, let us approximate the direction of the galactic centre (which is only $5.5$ degrees off the plane of the ecliptic)\footnote{The latitude $\beta$ and longitude $\lambda$ of the galactic centre in the standard ecliptic coordinate system are $\beta=-5.5^\circ$ and $\lambda=-93.2^\circ$ (see e.g. \citealt{allen}).} as being located in the plane of the orbit; consequently we choose $I=0$. In this case $\tilde{\omega}=\omega+\Omega$ is the relevant angle for the argument of the perihelion. The non-zero evolution equations then become
\begin{subequations}\label{avperteqsI0}
\begin{eqnarray}
\left\langle\frac{\ud e}{\ud t}\right\rangle &=& \frac{5 Q_2 e \sqrt{1-e^2}}{4n} \sin (2\tilde{\omega})\,,\\
\left\langle\frac{\ud \ell}{\ud t}\right\rangle &=& n - \frac{Q_2}{12 n}\Bigl[ 7 + 3 e^2 + 15 (1+e^2) \cos (2\tilde{\omega})\Bigr]\,,\\
\left\langle\frac{\ud \tilde{\omega}}{\ud t}\right\rangle &=& \frac{Q_2 \sqrt{1-e^2}}{4 n}\Bigl[ 1 + 5 \cos (2\tilde{\omega})\Bigr]\,.
\end{eqnarray}\end{subequations}
We recall that with our notation $\tilde{\omega}$ is the azimuthal angle between the direction of the perihelion and that of the galactic centre (approximated to lie in the orbital plane). Of particular interest is the secular precession of the perihelion $\langle\ud \tilde{\omega}/\ud t\rangle$ due to the quadrupole effect which we henceforth denote by\footnote{The result found by \cite{Milg09} for this effect [his Eqs.~(37)--(38)] looks more complicated, but can be simplified and is seen to be equivalent to ours.}
\begin{equation}\label{Delta2}
\Delta_2 = \frac{Q_2 \sqrt{1-e^2}}{4 n}\Bigl[ 1 + 5 \cos (2\tilde{\omega})\Bigr]\,.
\end{equation}
The precession is non-spherical, in the sense that it depends on the orientation of the orbit relative to the galactic centre through its dependence upon the perihelion's longitude $\tilde{\omega}$. The effect scales with the inverse of the orbital frequency $n=2\pi/P$ and therefore becomes more important for outer planets like Saturn than for inner planets like Mercury. This is in agreement with the fact that the quadrupole effect we are considering increases with the distance to the Sun (but of course will fall down when $r$ becomes appreciably comparable to $r_0$, see Fig.~\ref{f:qralpha1}). 

We have also computed the secular planetary precession induced by the octupole moment $Q_3$, for which the perturbation function reads with the same conventions
\begin{equation}\label{R3}
R = -\frac{1}{6} \,r^3 \,Q_3 \left(\cos^3\th - \frac{3}{5}\cos\th\right) = - \frac{Q_3}{30} x\,\Bigl(2x^2-3y^2-3z^2\Bigr)\,.
\end{equation}
Redoing the perturbation analysis we find the octupolar precession in the case $I=0$ as
\begin{equation}\label{Delta3}
\Delta_3 = \frac{Q_3 a \sqrt{1-e^2}}{32 e n}\Bigl[ 2 - 13 e^2 + 35 e^2 \cos (2\tilde{\omega})\Bigr]\cos \tilde{\omega}\,.
\end{equation}

\subsection{Numerical evaluation of the planetary precession}
\label{ss:precession}

We now compute numerically this effect for the various planets of the Solar System; the relevant orbital elements for the planets we used in this calculation are provided in Table~\ref{tab2}.
\begin{table*}
\begin{center}
\caption{Orbital elements of planets used in the computation of the abnormal precession rates $\Delta_2$ and $\Delta_3$. The longitude of the perihelion $\tilde{\omega}$ is defined here with respect to the direction of the external galactic field (assuming that the galactic centre lies within the plane of the ecliptic). Thus we pose $\tilde{\omega}=\tilde{\omega}_\text{Allen}-\lambda$, where $\lambda=-93.2^\circ$ is the longitude of the galactic centre and $\tilde{\omega}_\text{Allen}$ is given in Allen (1999).}
\label{tab2}
\begin{tabular}{lcccccccc}
\hline
	& Mercury & Venus & Earth & Mars & Jupiter & Saturn & Uranus & Neptune\\
\hline
	$a$ $[\text{AU}]$& $0.397$ & $0.723$ & $1.00$ & $1.523$ & $5.203$ & $9.537$ & $19.229$ & $30.069$ \\
	$e$ & $0.206$ & $0.007$ & $0.017$ & $0.093$ & $0.048$ & $0.054$ & $0.047$ & $0.009$ \\
	$P$ $[\text{yr}]$& $0.24$ & $0.62$ & $1.00$ & $1.88$ & $11.86$ & $29.46$ & $84.01$ & $164.8$ \\
	$\tilde{\omega}$ $[\text{deg}]$& $170.7$ & $224.7$ & $196.1$ & $69.2$ & $108.0$ & $185.6$ & $264.2$ & $138.2$ \\
\hline
\end{tabular}	
\end{center}
\end{table*}

\begin{table*}
\begin{center}
\caption{Results for the precession rates of planets $\Delta_2$ due to the quadrupole coefficient $Q_2$. We use the values for $Q_2$ corresponding to various MOND functions defined in Eqs.~\eqref{mu}-\eqref{e:def_mu_n} (see Table~\ref{tab1}). All results are given in \textit{milli}-arc-seconds per century ($\text{mas}/\text{cy}$).}
\label{tab3}
\begin{tabular}{ccccccccc}
\hline
\multicolumn{9}{|c|}{Quadrupolar precession rate $\Delta_2$ in $\text{mas}/\text{cy}$}\\
\hline
MOND function & Mercury & Venus & Earth & Mars & Jupiter & Saturn & Uranus & Neptune\\
\cline{2-9}
\hline
	$\mu_1(y)$ & $0.04$ & $0.02$ & $0.16$ & $-0.16$ & $-1.12$ & $5.39$ & $-10.14$ & $7.93$ \\
	$\mu_2(y)$ & $0.02$ & $0.01$ & $0.09$ & $-0.09$ & $-0.65$ & $3.12$ & $-5.87$ & $4.59$ \\
	$\mu_5(y)$ & $7\times 10^{-3}$ & $3\times 10^{-3}$ & $0.03$ & $-0.03$ & $-0.22$ & $1.05$ & $-1.97$ & $1.54$ \\
	$\mu_{20}(y)$ & $2\times 10^{-3}$ & $10^{-3}$ & $9\times 10^{-3}$ & $-9\times 10^{-3}$ & $-0.06$ & $0.3$ & $-0.56$ & $0.44$ \\
	$\mu_{\textrm{exp}}(y)$ & $0.03$ & $0.02$ & $0.13$ & $-0.13$ & $-0.88$ & $4.25$ & $-8.01$ & $6.26$ \\
	$\mu_{\textrm{TeVeS}}(y)$ & $0.05$ & $0.02$ & $0.17$ & $-0.17$ & $-1.21$ & $5.81$ & $-10.94$ & $8.56$ \\
\hline
\end{tabular}
\end{center}
\end{table*}
\begin{table*}
\begin{center}
\caption{Results for the precession rates of planets $\Delta_3$ due to the octupole coefficient $Q_3$. Beware of the results given here in \textit{micro}-arc-seconds per century ($\mu\text{as}/\text{cy}$).}
\label{tab4}
\begin{tabular}{ccccccccc}
\hline
\multicolumn{9}{|c|}{Octupolar precession rate $\Delta_3$ in $\mu\text{as}/\text{cy}$}\\
\hline
MOND function & Mercury & Venus & Earth & Mars & Jupiter & Saturn & Uranus & Neptune\\
\cline{2-9}
\hline
	$\mu_1(y)$ & $2\times 10^{-3}$ & $0.2$ & $0.2$ & $-0.03$ & $1.4$ & $19.5$ & $12.1$ & $1499$ \\
	$\mu_2(y)$ & $2\times 10^{-3}$ & $0.1$ & $0.2$ & $-0.03$ & $1.1$ & $15.1$ & $9.4$ & $1162$ \\
	$\mu_5(y)$ & $10^{-4}$ & $5\times 10^{-3}$ & $0.01$ & $-2\times 10^{-3}$ & $0.06$ & $0.8$ & $0.5$ & $61.2$ \\
	$\mu_{20}(y)$ & $5\times 10^{-5}$ & $2\times 10^{-3}$ & $5\times 10^{-3}$ & $-7\times 10^{-4}$ & $0.03$ & $0.4$ & $0.2$ & $28.7$ \\
	$\mu_{\textrm{exp}}(y)$ & $2\times 10^{-3}$ & $0.2$ & $0.2$ & $-0.03$ & $1.4$ & $19.5$ & $12.1$ & $1499$ \\
	$\mu_{\textrm{TeVeS}}(y)$ & $2\times 10^{-3}$ & $0.2$ & $0.2$ & $-0.03$ & $1.3$ & $17.9$ & $11.1$ & $1374$ \\
\hline
\end{tabular}
\end{center}
\end{table*}
Our numerical values for the quadrupole and octupole anomalies $\Delta_2$ and $\Delta_3$ are reported in Tables~\ref{tab3} and~\ref{tab4} respectively. As we see from Table~\ref{tab3} the quadrupolar precession $\Delta_2$ is in the range of the milli-arc-second per century which is not negligible. In particular it becomes interestingly large for the outer gaseous planets of the Solar System, essentially Saturn, Uranus and Neptune. The dependence on the choice of the MOND function $\mu$ is noticeable only for functions $\mu_n(y)$ defined by \eqref{e:def_mu_n} with large values of $n$, where the effect decreases by a factor $\sim 10$ between $n=2$ and $n=20$. However, the octupolar precession $\Delta_3$ in Table~\ref{tab4} is much smaller, being rather in the range of the \textit{micro}-arc-second per century. 

We give for comparison in Table~\ref{tab5} the best published postfit residuals for any possible supplementary precession of planetary orbits (after the relativistic precession has been duly taken into account), which have been obtained from global fits of the Solar System dynamics by \cite{pitjeva05} and \cite{fienga09}. In particular the INPOP planetary ephemerides by \cite{fienga09} use information from the combination of very accurate tracking data of spacecrafts orbiting different planets.
\begin{table*}
\begin{center}
\caption{We reproduce some published residuals for any supplementary orbital planetary precession. All results are given in milli-arc-seconds per century ($\text{mas}/\text{cy}$).} 
\label{tab5}
\begin{tabular}{lcccccccc}
\hline
\multicolumn{9}{|c|}{Postfit residuals for $\Delta=\langle\ud\tilde{\omega}/\ud t\rangle$ in $\text{mas}/\text{cy}$}\\
\hline
	Origin & Mercury & Venus & Earth & Mars & Jupiter & Saturn & Uranus & Neptune\\
\hline
	Pitjeva (2005) & $-3.6\pm 5$ & $-0.4\pm 0.5$ & $-0.2\pm 0.4$ & $0.1\pm 0.5$ & - & $-6\pm 2$ & - & - \\
	Fienga \textit{et al.} (2009) & $-10\pm 30$ & $-4\pm 6$ & $0\pm 0.016$ & $0\pm 0.2$ & $142\pm 156$ & $-10\pm 8$ & $0\pm 2\times10^4$ & $0\pm 2\times10^4$ \\
\hline
\end{tabular}
\label{known}
\end{center}
\end{table*}

From Table~\ref{tab5} we see that our results for $\Delta_2$ are smaller or much smaller than the published residuals except for the planets Earth, Mars and Saturn. The case of Saturn is interesting because the INPOP planetary ephemerides by \cite{fienga09} publish an offset with respect to the general relativistic value for the precession: namely they quote $-10\pm 8\,\text{mas}/\text{cy}$ as obtained from the Cassini tracking data and $200\pm 160\,\text{mas}/\text{cy}$ from the VEX data. The values we find for $\Delta_2$ are smaller but grossly within the range of these postfit residuals. However we find that $\Delta_2$ is positive for Saturn while the offset reported in \cite{fienga09} from Cassini data is negative. But note that this may not be relevant because the INPOP ephemerides are used by \cite{fienga09} to detect the presence of an eventual abnormal precession, not to adjust the value of that precession.

In the case of the Earth the INPOP ephemerides find a tight constraint $0\pm 0.016$ coming from the Jupiter VLBI data \citep{fienga09}. This constraint seems already to exclude most of our obtained values for $\Delta_2$, except for MOND functions of type $\mu_n$~\eqref{e:def_mu_n} with rather large values of $n$. In particular, one needs to take $n>8$ in order to have an EFE precession compatible with this constraint on the Earth orbit. On the other hand we note that for all the planets the octupolar precession rate $\Delta_3$ is very small and clearly completely negligible in current fits of the Solar System dynamics.

Thus it seems that in the case of the Earth the constraint from the Jupiter VLBI data is already quite tight, and it is difficult to accommodate our anomalous quadrupolar precession rate $\Delta_2$ for most choices of MOND function $\mu(y)$. However let us remark that the postfit residuals in Table~\ref{tab5} are obtained by adding by hands an excess of precession for the planets and looking for the tolerance of the data on this excess. But in order to really test the anomalous quadrupolar precession rate $\Delta_2$, one should consistently work in a MOND picture, i.e. consider also the other effects predicted by this theory, like the precession of the nodes, the variation of the eccentricity and the inclination, and so on. Then one should perform a global fit of all these effects to the data; it is likely that in this way the quantitative conclusions would be different.

On the other hand, as we have commented in the Introduction, it is non-trivial to know if testing MOND-like theories (including TeVeS and the model proposed by \citealt{BL08,BL09}) in the Solar System could invalidate those theories, if they represent only some phenomenological models describing the weak field regime of gravity and as such should not be extrapolated as they are in the strong field of the Solar System. 

\section{Conclusion} \label{s:conc}

In MONDian gravity, the influence of the external galactic gravitational field onto the orbits of Solar-system planets appears mainly through the presence of a quadrupolar perturbation to the gravitational potential, as it has been shown through a multipolar analysis of the solution of the MOND equation. This contribution has been computed using a spectral, highly accurate, numerical Poisson solver to obtain a solution of the equation~\eqref{e:MOND}. This numerical solution has been extensively tested and compared to known analytical formulas, and all the expected properties have been recovered. Using the most commonly used MOND coupling functions $\mu(y)$ and value of the MOND acceleration $a_0$, we have obtained quadrupole contributions that are compatible with those published by \cite{Milg09}. We have also extended this calculation to include octupole contributions. With the derivation of detailed perturbation equations, the influence of the quadrupole and octupole on the perihelion precession of planets have been quantified and compared to two sets of observations, in particular the INPOP planetary ephemerides by \cite{fienga09}. 

At first sight, these observational results seem to exclude the presence of a MOND-EFE induced quadrupole, for some choices of the MOND function $\mu$ and scale $a_0$. This is particularly true for the Earth orbit, where constraints from the Jupiter VLBI observations are very strong. However, for other choices of the MOND function the observational constraints on the precession of the Earth orbit are still compatible with the computed effect. These compatible MOND functions share the property of having a very sharp transition from the weak-field (modified) to the strong field (Newtonian) regime. Moreover one should be very cautious and keep in mind that:
\begin{itemize}
\item[1.] The observational constraints on the Solar-system orbits have been obtained by a global fit, using a fully relativistic first-post-Newtonian model. Taking only into account a particular MOND effect like precession without considering a fully-MONDian model is in principle incoherent since all the other effects should also be considered;
\item[2.] The MOND theory is a phenomenological approach trying to describe gravity in the weak field regime (where evidence for dark matter arises). It is therefore unclear if such theory can be applied to stronger gravitational fields and, in particular, in the Solar System. Conversely, the constraints obtained here in strong field may not be relevant for the weak-field case unless some more fundamental theory, valid in both regimes, is known.
\end{itemize}
In any case, further studies are to be done if one wants to obtain more stringent conclusions about constraints imposed by Solar-system observations onto MOND-like theories. More precise observations could also give valuable informations about an eventual EFE due to MOND theory and also restrict the number of possible MOND functions that are compatible with the observations.

\section*{Acknowledgments}
It is a pleasure to thank Christophe Le Poncin-Lafitte, Alexandre Le Tiec and Mordehai Milgrom for a careful reading of the manuscript and very useful comments.
\\

\appendix

\section{Asymptotic behaviour of the field}
\label{appA}

In this Appendix we prove that the asymptotic behaviour of the solution of the Bekenstein-Milgrom equation~\eqref{e:MOND} at large distances from an isolated matter source in the presence of an external field $\bm{g}_\text{e}$, is given by Eq.~\eqref{e:asym_check}. The proof has already been given by \cite{BekM84}; here we present a slightly different derivation. For simplicity we suppose that the source is spherically symmetric, so that the asymptotic form of the MOND potential is axisymmetric around the direction of the external field and of the type
\begin{equation}
  \label{Uf}
  U = g^i_\text{e}x^i + \frac{f(\th)}{r} + \mathcal{O} \left(
    \frac{1}{r^2} \right)\,,
\end{equation}
where $f(\th)$ is a function of the azimuthal angle $\th$ measured from the direction $\bm{e}$ of the external field. We naturally assume that the function $f(\th)$ is regular. With the ansatz \eqref{Uf} we obtain successively the leading order terms in the expansion of various quantities as
\begin{subequations}
\begin{align}
  \label{gif}
  g^i &= g^i_\text{e} + \frac{1}{r^2}\left[\left(-f + \frac{\cos\th}{\sin\th}f_\th\right)n^i - \frac{f_\th}{\sin\th} e^i\right] + \mathcal{O} \left(
    \frac{1}{r^3} \right)\,,\\
\mu &= \mu_\text{e}\left( 1 + \frac{\lambda_\text{e}}{g_\text{e} r^2}\Bigl[-\cos\th f - \sin\th f_\th\Bigr]\right) + \mathcal{O} \left(
    \frac{1}{r^3} \right)\,,\\
\mu g^i &= \mu_\text{e} \left(g^i_\text{e} + \frac{1}{r^2}\left[\left(- \lambda_\text{e} f \cos\th - \lambda_\text{e} f_\th \sin\th - \frac{f_\th}{\sin\th}\right) e^i \right. \right.\nonumber\\
&\left. \left.+ \left(-f + \frac{\cos\th}{\sin\th}f_\th\right)n^i \right]\right) + \mathcal{O} \left(\frac{1}{r^3} \right)\label{mugi}\,,
\end{align}\end{subequations}
where we denote $f_\th\equiv \partial f/\partial\th$, $\mu_\text{e} \equiv \mu(y_\text{e})$, $\lambda_\text{e} \equiv y_\text{e} \mu'_\text{e}/\mu_\text{e}$ (with $y_\text{e}=g_\text{e}/a_0$ and $\mu'_\text{e}= \ud\mu(y_\text{e})/\ud y_\text{e}$). Next, we impose that $\partial_i(\mu g^i)=0$ holds asymptotically far from a localised matter distribution. This, together with the fact that $f$ is a regular function of the azimuthal angle $\th$, yields the first-order differential equation
\begin{equation}
  \label{soleqdiff}
  \bigl(1 + \lambda_\text{e}\sin^2\th\bigr) f_\th + \lambda_\text{e}\sin\th\cos\th f = 0\,,
\end{equation}
whose solution is given in terms of a constant $K$ by
\begin{equation}
  \label{soleqdiff_sol}
  f = \frac{K}{\sqrt{1 + \lambda_\text{e}\sin^2\th}}\,.
\end{equation}
Finally we determine this constant by using the Gauss theorem \eqref{e:Gauss}, i.e.
\begin{equation}
  \label{e:Gausssphere}
  \oint \mu g^i r^2 n^i \textrm{d}\Omega = -4\pi G M, 
\end{equation}
where we choose a coordinate sphere since the result does not depend on the chosen surface at infinity. Inserting into \eqref{e:Gausssphere} the asymptotic expansion of the field given by \eqref{mugi}, together with the solution \eqref{soleqdiff_sol} found for $f$, we obtain
\begin{equation}
K = \frac{G M}{\mu_\text{e}}\,,
\end{equation}
which establishes the looked-for formula \eqref{e:asym_check}.

\section{Expression of the multipole expansion}\label{appB}

The result~\eqref{expu} can also be proved directly, without resorting to the computation of surface integrals of Sec.~\ref{s:multipoles}. We first expand $\delta u = u - u_\text{N}$ when $r\to 0$ using Taylor's formula as
\begin{equation}
  \label{deltau1}
\delta u = \sum_{l=0}^{+\infty}\frac{1}{l!} x^L (\partial_L \delta u)(\mathbf{0})\,.
\end{equation}
Here the derivative operator $\partial_L$ is \textit{a priori} non STF but we can decompose it into a sum of STF components using the formula (see the Appendix~\ref{appA} of \citealt{BD86} for a compendium of useful formulas)
\begin{equation}
  \label{STFformula}
\partial_L = \sum_{k=0}^{[l/2]} a_k^l\,\delta_{(2K}\,\hat{\partial}_{L-2K)}\,\Delta^k\,,
\end{equation}
where $k$ ranges from 0 to the integer part of $l/2$ denoted $[l/2]$, where $\delta_{2K}$ denotes a product of $k$ Kronecker symbols carrying $2k$ indices chosen among those of $L = i_1 \cdots i_l$, where the parenthesis indicate full symmetrisation over the $l$ indices $i_1 \cdots i_l$, where $\Delta\equiv\partial_{ii}$ denotes the ordinary Laplacian, and where the coefficients depending on $l$ and $k$ are given by
\begin{equation}
  \label{alk}
a_k^l = \frac{l!}{(l-2k)!}\frac{(2l-4k+1)!!}{(2k)!!(2l-2k+1)!!}\,.
\end{equation}
Using now the fact that $\Delta \delta u=0$ in a neighbourhood of the origin ($r\leqslant\varepsilon$) we see that only the term $k=0$ will survive in the latter sum, and since $a_{k=0}^l=1$ we conclude that we can replace in fact the derivative operator in $(\partial_L \delta u)(\mathbf{0})$ by its STF projection $\hat{\partial}_L$, hence
\begin{equation}
  \label{deltau2}
\delta u = \sum_{l=0}^{+\infty}\frac{1}{l!} x^L (\hat{\partial}_L \delta u)(\mathbf{0})\,.
\end{equation}
Next, we prove that the contribution from the Newtonian potential $u_\text{N}$ in \eqref{deltau2} is purely monopolar. This comes from our assumption that the density of ordinary matter $\rho$ in the Sun is both regular (i.e. $C^\infty$) and spherically symmetric, i.e. depending only on $r$ (neglecting the back-reaction of the field on its source). As we shall now prove, this means that the Taylor expansion of $\rho(r)$ in spherical symmetry must contain only even powers of $r$. The density being regular its expansion when $r\to 0$ is given by Taylor's formula as
\begin{equation}
\rho = \sum_{l=0}^{+\infty}\frac{1}{l!} x^L (\partial_L \rho)(\mathbf{0})\,.
\label{e:Taylor_rho}
\end{equation}
Next we again use the formula \eqref{STFformula} to decompose \eqref{e:Taylor_rho} into a sum of STF pieces. Because $\rho$ is spherically symmetric all these STF pieces must be zero except the monopolar contribution; hence only the terms for which $l=2k$ survive in the latter decomposition. Using the value of the coefficient $a_k^{2k}=1/(2k+1)$ we are then left with
\begin{equation}
\rho = \sum_{k=0}^{+\infty}\frac{r^{2k}}{(2k+1)!} \,(\Delta^k\rho)(\mathbf{0})\,.
\label{e:Taylor_rho_spher}
\end{equation}
As we see $\rho(r)$ admits an expansion containing only even powers of $r$. Now the spherically-symmetric Newtonian potential $u_\text{N}$ will also admit an expansion only in even powers of $r$, so we write accordingly $u_\text{N}(r)=f(r^2)$. Acting on a function of $r^2$ the STF derivative operator gives $\hat{\partial}_Lf(r^2)=\hat{n}^L(2r)^l f^{(l)}(r^2)$ where $f^{(l)}$ denotes the $l$-th derivative of this function (see \citealt{BD86}). Hence we find that $\hat{\partial}_Lu_\text{N}$ is a perfectly regular function and we conclude that $(\hat{\partial}_Lu_\text{N})(\mathbf{0})=0$ except for the monopole case $l=0$. We can therefore replace $(\hat{\partial}_L \delta u)(\mathbf{0})$ in \eqref{deltau2} by $(\hat{\partial}_L u)(\mathbf{0})$ for any $l\geqslant 1$ and we have finally established the result
\begin{equation}
  \label{deltau3}
\delta u = - u_\text{N}(\mathbf{0}) + \sum_{l=0}^{+\infty}\frac{1}{l!} x^L (\hat{\partial}_L u)(\mathbf{0})\,.
\end{equation}
which is the same as in Eq.~\eqref{expu}.

\label{lastpage}
\end{document}